\documentclass[journal=acsodf,manuscript=article]{achemso}
\usepackage{graphicx,subcaption,amsmath, textcomp,gensymb,esvect,xcolor,xr}

\usepackage{array}
\usepackage{multirow}

\setkeys{acs}{maxauthors=30, etalmode=truncate}

   \makeatletter
\newcommand*{\addFileDependency}[1]{% argument=file name and extension
  \typeout{(#1)}
  \@addtofilelist{#1}
  \IfFileExists{#1}{}{\typeout{No file #1.}}
}
\makeatother

\usepackage[version=3]{mhchem}

\usepackage[normalem]{ulem}

\newenvironment{competing interests}% environment name
{% begin code
  \par\vspace{\baselineskip} \noindent
  \begin{Large}\textbf{Competing Interests}\end{Large} 
  \par \noindent\ignorespaces
}%% end code

\newenvironment{data availability}% environment name
{% begin code
  \par\vspace{\baselineskip}\noindent
  \begin{Large}\textbf{Data Availability}\end{Large}
  \par \noindent\ignorespaces
}%% end code

\newenvironment{author contribution}% environment name
{% begin code
  \par\vspace{\baselineskip}\noindent
  \begin{Large}{\textbf{Author Contribution}} \end{Large}
  \par \noindent\ignorespaces
}%% end code

\newenvironment{tableofc}% environment name
{% begin code
  \par\vspace{\baselineskip}\noindent
  \begin{Large}{\textbf{For Table of Contents Only}} \end{Large}
  \par \noindent\ignorespaces
}%% end code

\author{Sherifdeen O. Bolarinwa}
\affiliation{Department of Physics, Faculty of Science, King AbdulAziz University, B. O. Box 80203, Jeddah 21589, Saudi Arabia}
\author{Shahid Sattar}
\email{shahid.sattar@lnu.se}
\affiliation{Department of Physics and Electrical Engineering, Linnaeus University, SE-39231 Kalmar, Sweden}
\author{Abdullah A. AlShaikhi}
\email{aalshaikhi@kau.edu.sa}
\affiliation{Department of Physics, Faculty of Science, King AbdulAziz University, B. O. Box 80203, Jeddah 21589, Saudi Arabia}

\title{Superior Gas Sensing Properties of $\beta$-In$_2$Se$_3$: A First-Principles Investigation}


\begin{document}

\begin{abstract}
Using first-principles calculations, we report structural and electronic properties of CO, NO$_2$ and NO molecular adsorption on $\beta$-In$_2$Se$_3$ in comparison to a previous study on $\alpha$-phase. Analysis and comparison of adsorption energies and extent of charge transfer indicates $\beta$-In$_2$Se$_3$ to be selective in detecting gas molecules. We found NO molecules acting as charge donor whereas CO and NO$_2$ molecules as charge acceptors, respectively, experiencing physisorption in all cases. Owing to enhanced adsorption, faster desorption and improved selectivity of the gas molecules discussed in detail, we conclude $\beta$-In$_2$Se$_3$ to be a superior gas sensing material ideal for chemoresistive-type gas sensing applications.

\end{abstract}
\newpage

\section{Introduction}

The need to identify gas leakage, toxic gases, and organic vapours for human and environmental safety is fundamental to develop next-generation sensing technologies. Gases that are by-products of our day-to-day activities (e.g., CO, NO2, and NO etc.) and toxic even at lower concentrations demand great attention. Different materials are thus employed in a bid to identify toxic gases which includes conducting polymers \cite{janata2003conducting,miasik1986conducting,virji2004polyaniline}, carbon nanotubes (CNTs) \cite{li2003carbon,wang2009review}, and semiconducting metal oxides \cite{kanan2009semiconducting,sun2012metal} to list a few. Conducting polymers are easily processed, but effects of humidity and degradation hamper their applicability \cite{bai2007gas,yoon2013current,cheah1998ordering}. On the other hand, metal oxides show prospects in sensing molecules, however, high operating temperature, large power consumption, and low selectivity have been a major issue in their commercial usage \cite{fine2010metal,ponzoni2017metal,wang2010metal}. Owing to these obvious drawbacks, researchers have intensified efforts in exploring potential new materials that can efficiently detect gases at room temperature and standard environmental conditions while retaining high selectivity and sensibilities.

The successful synthesis of graphene \cite{novoselov2004electric} earlier this century has birthed an extensive and increasing research attention on two dimensional (2D) materials. Since then, 2D materials have been explored for countless applications (such as Li-ion batteries \cite{chang2011cysteine}, electrocatalysis \cite{kibsgaard2010comparative}, photon-emitting \cite{lee2012synthesis} devices to name a few). Their inimitable geometry \cite{wang2012electronics}, large surface-to-volume ratio \cite{novoselov20162d} with strong surface reactivity \cite{gupta2015recent} and possibility to achieve thickness-modulated sensing characteristics \cite{zhang2015experimental} make 2D materials highly suitable for gas sensing applications. These remarkable features have led to an increased use of 2D materials for gas-sensing applications and there exist several theoretical and experimental reports showing good sensing abilities with high response \cite{chu2018highly,wang2018first,zhang2019facile}.

Indium selenide (In$_2$Se$_3$) is a promising 2D semiconductor belonging to the III$_2$-VI$_3$ chalcogenides covering a broad range of light absorption \cite{feng2018high}, having tunable bandgap \cite{jacobs2014extraordinary,kang2014thermally}, and phase change properties \cite{kang2014thermally,feng2018phase}. It is a polymorphic non-transition metal chalcogenide that crystallizes in five distinct polymorphs namely $\alpha$, $\beta$, $\gamma$, $\delta$ and $\kappa$ depending on growth temperature and pressure \cite{zhou2015controlled,tao2013crystalline,li2011thermal,popovic1979revised,vilaplana2018experimental,debbichi2015two}. The $\alpha$- and $\beta$-phases experience van der Waals interactions \cite{zhou2015controlled,zheng2017self} making them suitable for several applications including phase-change memory \cite{yu2007indium}, e-skin applications \cite{feng2016sensitive}, optoelectronics \cite{jacobs2014extraordinary,zhai2010fabrication}, and photodetection \cite{quereda2016strong} to list a few. While both phases possess similar geometries, they exhibit significant difference in electronic properties \cite{debbichi2015two}. The electronic band structure of $\beta$-In$_2$Se$_3$ is reported to have stronger sensitivity to externally applied electric fields compared to the $\alpha$-phase \cite{tao2013crystalline, debbichi2015two}. However, while the gas sensing applications of $\alpha$-In$_2$Se$_3$ have been reported in literature \cite{xie2018functionalization}, the gas sensing abilities of atomically thin $\beta$-In$_2$Se$_3$ have so far remain lacking. In addition, the existing report on $\alpha$-In$_2$Se$_3$ does not take into account the van der Waals dispersion corrections for the molecular adsorptions. It is therefore in this study we investigate gas sensing properties of $\beta$-In$_2$Se$_3$ for CO, NO$_2$, and NO molecules by using first-principles calculations incorporating van der Waals dispersion corrections. Various atomic sites were carefully inspected for molecular absorption to reveal the most favourable structural configurations, for which electronic band structures and density-of-states are calculated. We found the NO molecule to have the highest binding energy whereas Bader charge analysis shows significant charge transfer for NO$_2$ molecule compared to others. Similarly, we observed that the NO$_2$ physisorption gave rise to mid-gap states whereas NO molecules strongly perturb region close to the conduction band minimum. Our analysis and comparison to the $\alpha$-phase shows that $\beta$-In$_2$Se$_3$ is a promising material for chemoresistive based gas sensing applications.

\section{Computational Method}

We have performed first-principles calculations based on density functional theory (DFT) \cite{hohenberg1964inhomogeneous,kohn1965self} as implemented in the Quantum Espresso package \cite{giannozzi2009quantum,giannozzi2017advanced}. To describe exchange and correlation effects, we used generalized gradient approximations (GGA) in the Perdew-Burke-Ernzerhof parameterization (PBE) \cite{perdew1996generalized}. The projected augmented wave (PAW) method of DFT with plane-wave cut-off energy truncated at 45 Ry was used \cite{blochl1994projector}. For structural relaxation (self-consistent calculations), we used $2\times2\times 1$ ($4\times4\times 1$) k-mesh, respectively, whereas a dense $8\times8\times 1$ k-mesh was used for density-of-states (DOS) calculations. To avoid spurious interactions between the periodic images and efficiently model the adsorbents, a $3\times3\times 1$ supercell of $\beta$-In$_2$Se$_3$ with 15 \AA\,vacuum was constructed. In the iterative solution of the Kohn-Sham equations, an energy convergence of 10$^{-4}$ Ry and a force convergence of 10$^{-3}$ Ry/Bohr was achieved. We used Grimme’s DFT-D3 approach to describe the interlayer vdW interactions \cite{grimme2010consistent,grimme2011effect}. Bader charge transfer was performed using the method described in Ref. \cite{henkelman2006fast}.

\section{Results and Discussion}

\begin{figure}[t]
\includegraphics[width=0.99\textwidth]{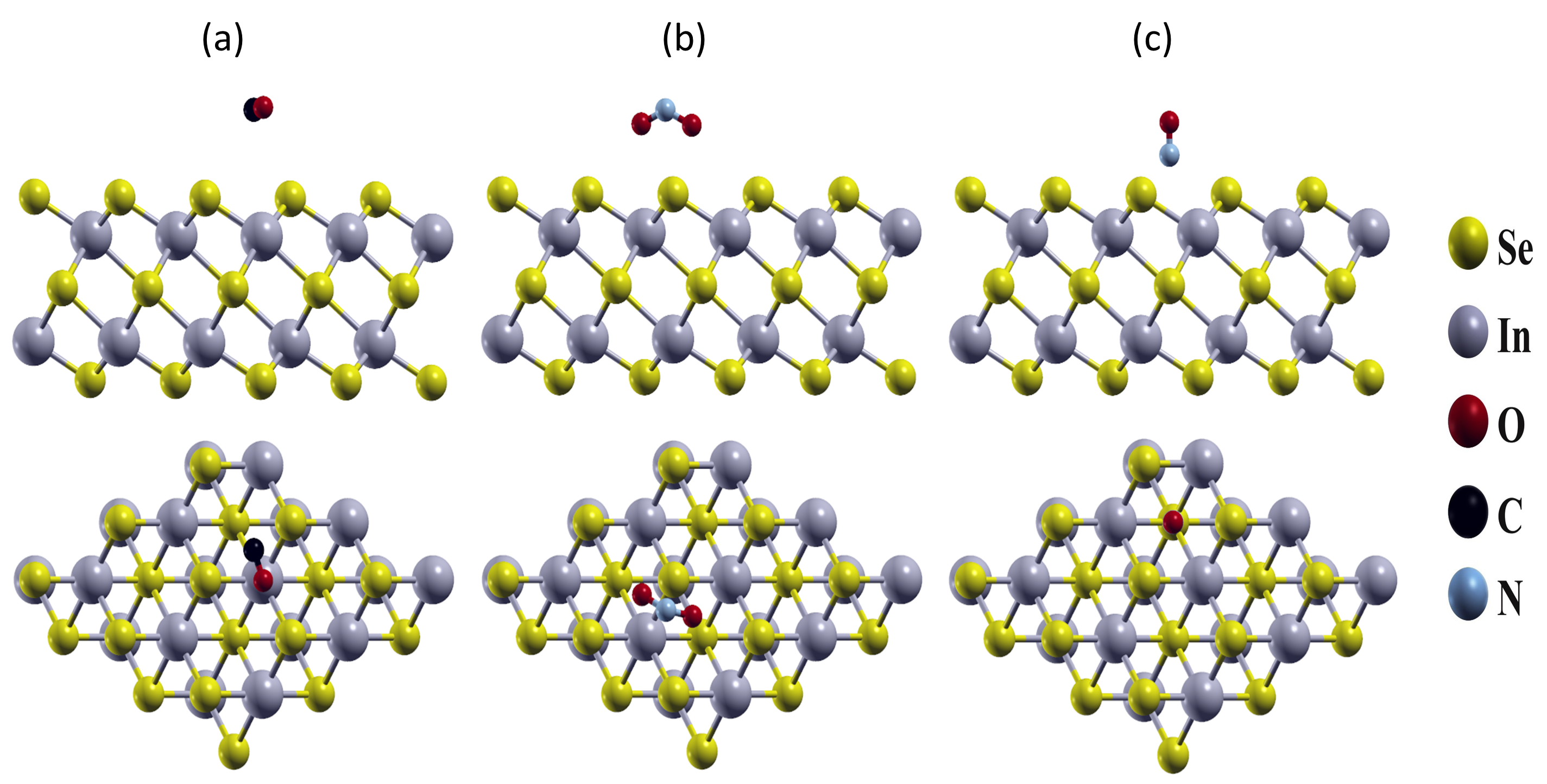}
\caption{Top and side views of the energetically favourable adsorption sites of (a) CO (b) NO$_2$ and (c) NO on $\beta$-In$_2$Se$_3$, respectively.}
\label{fig:fig1}
\end{figure}

The optimized lattice constant of $\beta$-In$_2$Se$_3$ turns out to be 4.03 \AA\,which is in good agreement to the experimental value of 4.025 \cite{li2018large}. It consists of five covalently bonded atomic sheets known as quintuple layers (QL) that are vertically stacked in the sequence Se-In-Se-In-Se atoms. Each stratum in the QL takes the form of a well-aligned triangular lattice which is also characterized by a surface projected hexagonal void. As a precondition to investigate gas sensing, we first identify the energetically favourable adsorption sites for CO, NO$_2$, and NO molecules on pristine $\beta$-In$_2$Se$_3$. Four anchoring sites have been considered with the centre of mass of the gas molecule initially positioned on top of multiple sites. These include (1) on top of the Se atom (T$_{\text{Se}}$), (2) on top of the In atom (T$_{\text{In}}$), (3) on top of the hexagonal ring (T$_{\text{HR}}$), and (4) on top of the In-Se bond (T$_{\text{B}}$). For each of these cases, we also examine different molecular orientations. Figure \ref{fig:fig1}(a-c) shows top and side views of the most favourable configurations for each of the molecules considered in this study.

Figure \ref{fig:fig1}a shows top and side views of the most stable configuration of CO molecule lying above the In atom parallel to the monolayer which is in line with existing study on InSe \cite{ma2017first}. After structural relaxation, the C-O bond length remains unchanged to 1.14 \AA\,, whereas the interlayer distance turns out to be 3.21 \AA\,clearly demonstrating the case of physisorption. The adsorption energy was calculated using equation (\ref{equation:equation1}), 
\begin{equation}
E_{\text{ads}} = E_{\text{In$_2$Se$_3$+M}} - E_{\text{In$_2$Se$_3$}} - E_{\text{M}},
\label{equation:equation1}
\end{equation}
where E$_{\text{In$_2$Se$_3$+M}}$, E$_{\text{In$_2$Se$_3$}}$, and E$_{\text{M}}$ corresponds to the total energies of $\beta$-In$_2$Se$_3$ with the adsorbed molecule, the pristine $\beta$-In$_2$Se$_3$ detached molecule, respectively. It follows from the above formula that a negative value of \textcolor{red}{-0.109 eV} is obtained implying an exothermic reaction, resembling the case of CO adsorption on graphene \cite{zarei2019conformational}.

\begin{table}[!htb]
\caption{Favourable atomic sites, interlayer distance between $\alpha$- or $\beta$-In$_2$Se$_3$ and molecule, adsorption energies $E_{\text{ads}}$, and magnitude of charge transfer ($\Delta$Q). \textcolor{red}{Data for the $\alpha$-In$_2$Se$_3$ was obtained from Ref. \cite{xie2018functionalization}.}} % title of Table

\begin{tabular}{|l|c|c|c|c|c|c|c|c|}
\hline\hline
\multirow{2}{*}{Molecule} & \multicolumn{2}{c|}{Site} & \multicolumn{2}{c|}{Distance (\AA)} & \multicolumn{2}{c|}{E$_{ads}$ (eV)} & \multicolumn{2}{c|}{$\Delta$ Q (e)} \\ \cline{2-9} 
                          & $\beta$        & $\alpha$ \cite{xie2018functionalization}        & $\beta$           & $\alpha$ \cite{xie2018functionalization}            & $\beta$           & $\alpha$ \cite{xie2018functionalization}              & $\beta$            & $\alpha$ \cite{xie2018functionalization}            \\ \hline
CO                        & T$_{\text{In}}$            & -        & 3.21              & -             & -0.109            & -               & 0.02         & -            \\ \hline
NO$_2$                       & T$_\text{B}$             & Se       & 2.97              & 3.57          & -0.190            & 0.058           & 0.14         & 0.081        \\ \hline
NO                        & T$_{\text{HR}}$            & Se       & 2.55              & 2.65          & -0.381            & 0.208           & 0.10         & 0.054        \\ \hline
\end{tabular}
\label{table:table1}
\end{table}

For the case of NO$_2$ molecule, the minimum energy configuration is shown in Figure \ref{fig:fig1}b, where the molecule prefers to sit at the top of the In-Se bond (bridge site) with the oxygen atoms pointing towards $\beta$-In$_2$Se$_3$. The N-O bond length was found to be 1.23 \AA\,with a reduced angle of 127.72$^{\circ}$ compared to 133.08$^{\circ}$ for the detached molecule. We observed an interlayer distance of 2.97 \AA\,with an adsorption energy of \textcolor{red}{-0.190 eV} alike the case on $\alpha$-In$_2$Se$_3$\cite{xie2018functionalization} and InSe \cite{ma2017first}. Continuing to the next case, we found that NO molecule favours adsorption at the centre of the hexagonal ring (T$_{\text{HR}}$) with a bond length of 1.15 \AA. It is relaxed perpendicularly such that the N atom is pointing towards $\beta$-In$_2$Se$_3$. The interlayer distance between the adsorbed NO molecule and $\beta$-In$_2$Se$_3$ was calculated to be 2.55 \AA\,, with a large adsorption energy of \textcolor{red}{-0.381 eV}. We notice that this configuration is also in line with those of NO adsorption on $\alpha$-In$_2$Se$_3$ \cite{xie2018functionalization} and InSe \cite{ma2017first} albeit with higher adsorption energy. It is also pertinent to mention that when compared with the adsorption behaviour on $\alpha$-In$_2$Se$_3$\cite{xie2018functionalization} our obtained adsorption energies are adequately large to inhibit room-temperature thermal fluctuations (see Table \ref{table:table1}) especially in the case of NO$_2$ on $\alpha$-In$_2$Se$_3$ where a weak physisorption energy of \textcolor{red}{-0.058 eV} was reported. In addition, our adsorption energy for NO and NO$_2$ molecules is indicative of a better adsorption/desorption recovery process on $\beta$-In$_2$Se$_3$ as compared to the existing adsorption case on $\alpha$-In$_2$Se$_3$. This inference takes a cue from the well known Sabatier's principle.\cite{karoyo2019surfactant} While we find the adsoprtion case of NO and NO$_2$ molecules on both phases to exhibits physisorption with similar molecular configurations, they are however distinct with respect to their favorable adsorption sites. No$_2$ molecule favors top of the In-Se bond (top of Se atom) while the NO molecule is located on top of the hexagonal ring (top of Se atom) for the $\beta$-In$_2$Se$_3$ ($\alpha$-In$_2$Se$_3$) phases.

\begin{figure}[!t]
\includegraphics[width=0.8\textwidth]{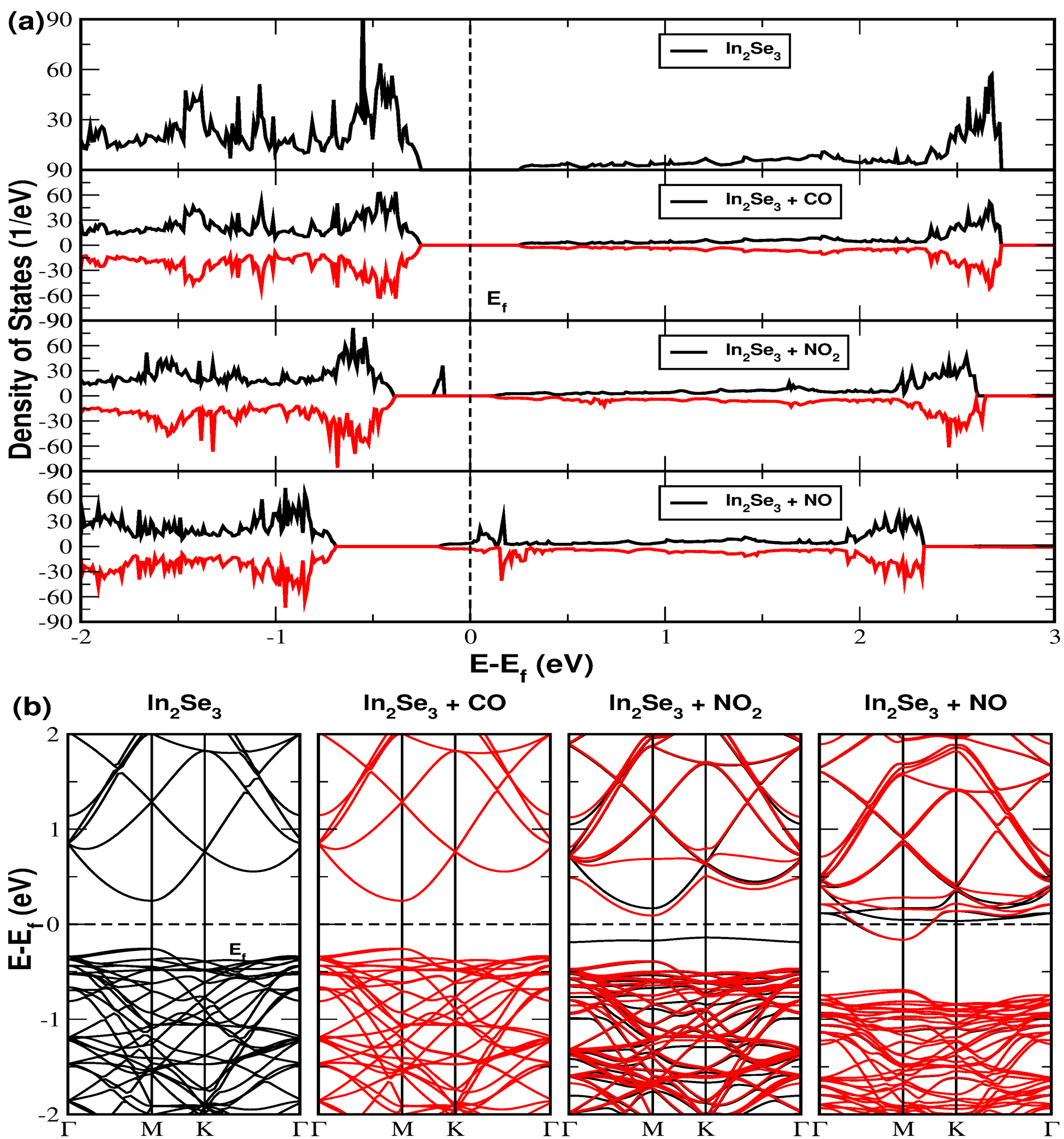}
\caption{(a) Density-of-states (DOS) for pristine and molecular adsorbed $\beta$-In$_2$Se$_3$. (b) Electronic band structures for pristine and molecular adsorbed $\beta$-In$_2$Se$_3$.}
\label{fig:fig2}
\end{figure}

Figure \ref{fig:fig2}a shows the electronic density-of-states (DOS) for pristine and molecular adsorbed $\beta$-In$_2$Se$_3$. For pristine $\beta$-In$_2$Se$_3$, we found a bandgap of 0.55 eV (at the level of GGA-PBE) for monolayer $\beta$-In$_2$Se$_3$ which is less than the experimental value of 1.55 eV \cite{almeida2017colloidal} but matches to the value 0.49 eV of an existing theoretical study \cite{li2018large}. Comparative analysis of pristine DOS to that of the CO molecular DOS reveals negligible effects on the electronic dispersion. States around the Fermi level remain unchanged without any orbital hybridization, however, there exists molecular contributions deep in the valence and conduction bands which do not play any role in sensing applications.

On the other hand, the DOS plot for NO$_2$ on $\beta$-In$_2$Se$_3$ indicates a significant contribution from the adsorbent with a noticeable downward shift of the \textcolor{red}{conduction band minima} (CBM) with respect to the Fermi level. We observe impurity state above the \textcolor{red}{valence band maxima} (VBM) and in the middle of the gap arising from the p-orbital of the adsorbates. A localized magnetic moment of 1.0 $\mu_B$ is also induced due to unpaired NO$_2$ electrons which results in spin-up and spin-down states falling at different energies.
Looking at the DOS plot of NO molecule, we observe appreciable changes with a downward shift of the CBM resulting in electron doping. We can infer the inducement of magnetic properties from the spin-polarized plots shown in Figure \ref{fig:fig2}a. Such property changes are useful in the sensing of NO molecules. This result corroborates those of NO molecule on $\alpha$-In$_2$Se$_3$\cite{vilaplana2018experimental} and InSe \cite{ma2017first}. Such alterations in electronic properties for the NO$_2$ and NO absorbent play a significant role in gas sensing by charge transfer based sensors. It is worthy of mentioning that the unpaired electrons in NO and NO$_2$ account for the formation of the localized states which consequently gave rise to the spin polarization effects. Similarly, the DOS plot indicates that NO$_2$ and NO has an enhanced electron transfer with the monolayer when compared to the case of CO molecule. We also plot the electronic band structure in Figure \ref{fig:fig2}b for the pristine and molecular adsorbed $\beta$-In$_2$Se$_3$. Evidently, no visible change is observed for the CO adsorption, whereas, for NO$_2$, a localized state is manifested above the VBM. On the other hand, for NO molecular adsorption, the band structure shows a downward shift in the CBM when compared to the pristine case.

\begin{figure}[!t]
\includegraphics[width=0.95\textwidth]{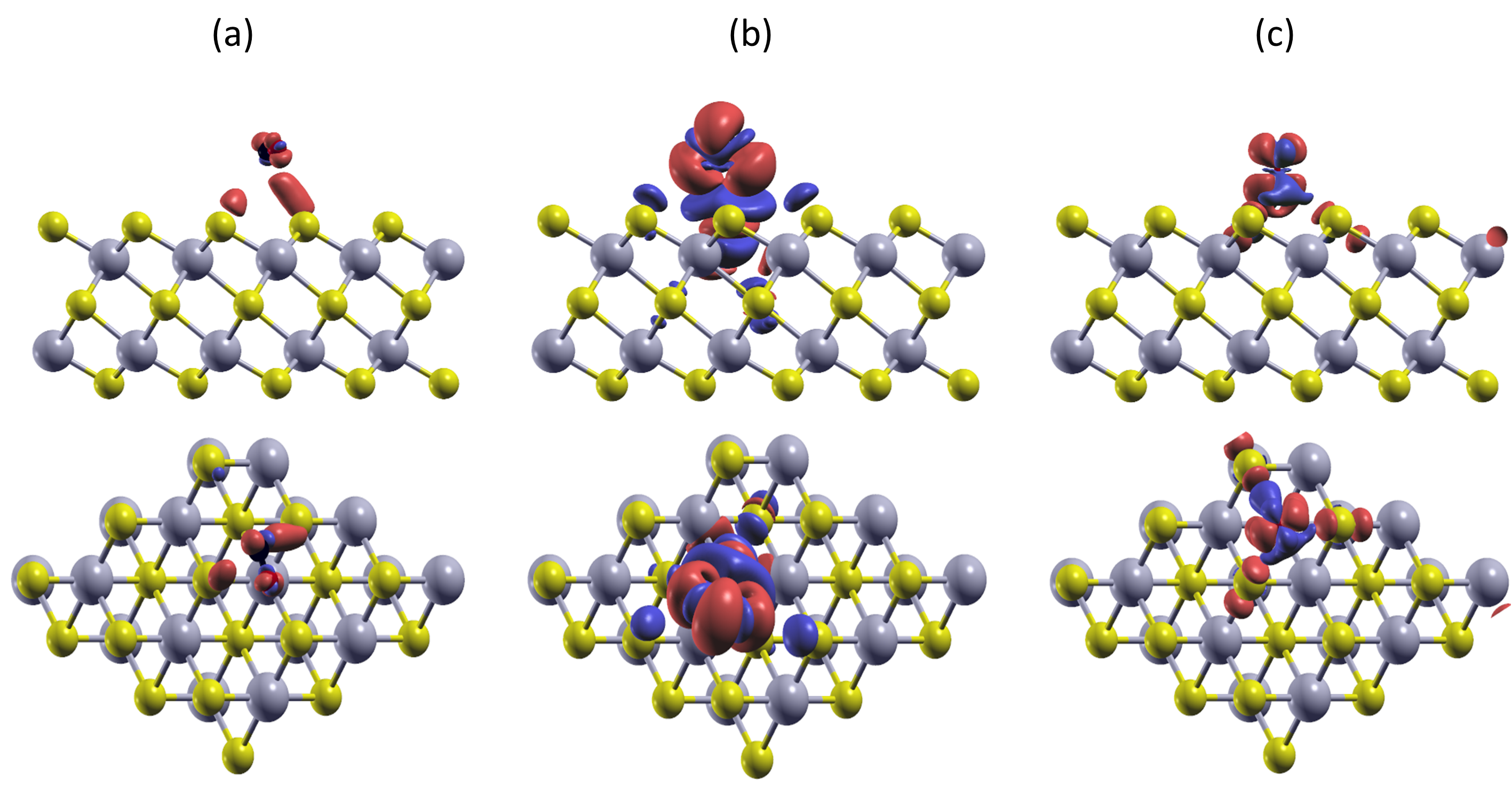}
\caption{ Top and side views of the charge density difference (CDD) plot for (a) CO (b) NO$_2$ and (c) NO molecules on $\beta$-In$_2$Se$_3$, respectively. Red and blue colors indicate charge accumulation and depletion, respectively. The isosurface value was set at $2\times10^{-3}$ eV/\AA$^{-3}$.}
\label{fig:fig3}
\end{figure}

To better understand charge transfer between $\beta$-In$_2$Se$_3$ and adsorbents, we next plot the charge density difference (CDD) in Figure \ref{fig:fig3} using equation $\Delta\rho=\rho_{\text{In$_2$Se$_3$+M}}-\rho_{\text{In$_2$Se$_3$}} -\rho_{\text{M}}$, whereas $\rho_{\text{In$_2$Se$_3$+M}}$, $\rho_{\text{In$_2$Se$_3$}}$, and $\rho_{\text{M}}$ represent charge densities for the monolayer with the adsorbed molecule, the pristine monolayer, and detached molecules, respectively. As shown in Figure \ref{fig:fig3}a, adsorption of CO results in charge redistribution within the molecule. Looking at the charge depletion from C atom that is accumulated on the O atom, we notice charge depletion from the Se atom tending towards the CO molecule. Using the Bader charge analysis, we found that the CO molecule accepts 0.02 e from the monolayer underneath confirming CO molecule to be a weak acceptor of electrons on $\beta$-In$_2$Se$_3$ as it were on InSe \cite{ma2017first}. The CDD for the adsorption of NO$_2$ on monolayer is shown in Figure \ref{fig:fig3}b. Evidently, NO$_2$ induces charge redistribution in the region between the molecule and the monolayer surface. The charge accumulation can be found around the molecule while the depletion is occurring mainly on the Se atom in the vicinity of the molecule, thereby indicating that NO$_2$ obtains charge from the monolayer. According to Bader charge analysis, NO$_2$ molecule accepts 0.14 e from $\beta$-In$_2$Se$_3$.

However, the converse was observed for the NO molecule adsorption on $\beta$-In$_2$Se$_3$ (see Figure \ref{fig:fig3}c). Charge depletion from the NO molecule is accumulated on monolayer surface (shown in red) which is corroborated by the Bader charge analysis showing a charge transfer of 0.1 e. We note here that the difference in the amount of charge transfer for each molecule could be indicative of the selectivity of $\beta$-In$_2$Se$_3$ to CO, NO$_2$, and NO molecules. When benchmarked with the existing study on $\alpha$-In$_2$Se$_3$, we observed that the charge transfer mechanism on both phases exhibited similar pattern with NO$_2$ as an acceptor and NO as a donor. However, the magnitude of the charge transfer to and from $\beta$-In$_2$Se$_3$ is almost double that of $\alpha$-In$_2$Se$_3$ \cite{xie2018functionalization}. It suggests that $\beta$-In$_2$Se$_3$ would sense better at lower concentrations of NO$_2$ and NO molecules compared to the $\alpha$-In$_2$Se$_3$ corroborating our earlier analysis.

\section{Conclusion}

Using first-principles calculations, we investigate structural and electronic properties of $\beta$-In$_2$Se$_3$ with the adsorption of CO, NO$_2$, and NO molecules. Experiencing physisorption in all cases, we found that $\beta$-In$_2$Se$_3$ would be selective in gas detection as CO and NO$_2$ acts as electron acceptors with 0.02 e and 0.14 e obtained respectively from the monolayer while NO acts as an electron donor by donating 0.02 e to the monolayer. In addition, we observed that CO had no significant effect on the electronic properties of the monolayer while NO$_2$ and NO introduced interstitial impurity states. It follows straight forwardly from the foregoing that $\beta$-In$_2$Se$_3$ from amongst the tested gases would sense best the NO molecules compared to other studied cases. In addition, our results suggest enhanced adsorption, faster desorption and improved selectivity of the gas molecules thereby making $\beta$-In$_2$Se$_3$ a superior gas sensing material compared to its $\alpha$-phase. Hence, our first-principles findings suggest that $\beta$-In$_2$Se$_3$ is a promising material in the fabrication of gas sensors built on the charge transfer mechanisms.

\section{Acknowledgement}
The computational resource for this calculation was provided by the High Performance Computing Centre at King Abdulaziz University (Aziz Supercomputer).

\begin{competing interests}
The Authors declare no competing financial or non-financial interests.
\end{competing interests}

\begin{data availability}
The data that support the findings of this study are available from the corresponding
author upon reasonable request.
\end{data availability}

\begin{author contribution}
S. O. Bolarinwa performed the calculations, S. Sattar and A. A. AlShaikhi analysed the results.  All authors reviewed the manuscript.\\ Corresponding Author:\\ Shahid Sattar, Email: shahid.sattar@lnu.se \\ Abdullah A. AlShaikhi, Email: aalshaikhi@kau.edu.sa,
\end{author contribution}

\bibliography{main.bib}

\begin{tableofc}
\\ \vspace{15mm}
\includegraphics{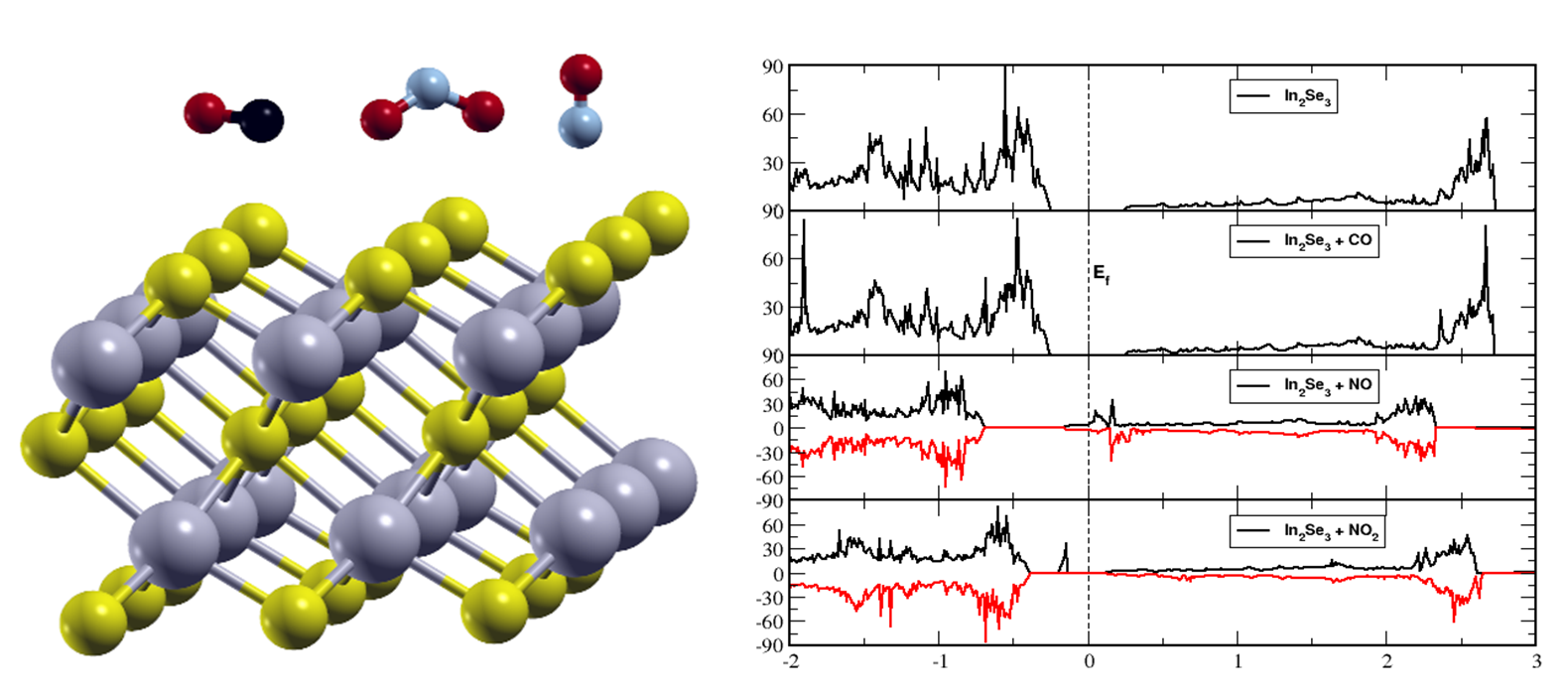}
\end{tableofc}

\end{document}